\documentclass[hidelinks,11pt]{article}

\usepackage{multirow}
\usepackage{amsmath}
\usepackage{amsfonts}
\usepackage{amsmath}
\usepackage{amssymb}  
\usepackage{graphicx}
\usepackage{caption}
\usepackage{mathrsfs}
\usepackage{upgreek}
\usepackage{subfig}
\usepackage{booktabs}
\usepackage{authblk}
\usepackage{stmaryrd}

\newcommand{\docircint}[2]{%
  \ifx#1\displaystyle
    \displaycircint
  \else
    \normalcircint{#1}%
  \fi
}
\newcommand{\displaycircint}{\displaystyle\mathsf{c}\mkern-25mu}
\newcommand{\normalcircint}[1]{%
  \smallerc{#1}\ifx#1\textstyle\mkern-9mu\else\mkern-8.2mu\fi
}
\newcommand{\smallerc}[1]{%
  \vcenter{\hbox{$\ifx#1\textstyle\scriptstyle\else\scriptscriptstyle\fi\mathsf{c}$}}%
}

\usepackage{color}
\newcommand{\edit}[1]{#1}
\newcommand{\secref}[1]{Sec.~\ref{#1}}

\renewcommand{\eqref}[1]{(\ref{#1})}
\newcommand{\citeasnoun}[1]{\cite{#1}}
\newcommand{\citeasnouns}[1]{\cite{#1}}
\renewcommand{\vec}[1]{\mathbf{#1}}

\title{Inverse design of large-area metasurfaces}
\author[1,2]{Rapha\"el~Pestourie\thanks{pestourie@g.harvard.edu}}
\author[1,3]{Carlos~P\'erez-Arancibia}
\author[1]{Zin~Lin}
\author[1]{Wonseok~Shin}
\author[2]{Federico~Capasso}
\author[1]{Steven~G.~Johnson}
\affil[1]{\small{Department of Mathematics, Massachusetts Institute of Technology, Cambridge, MA 02139, USA}}
\affil[2]{\small{Harvard John A. Paulson School of Engineering and Applied Sciences, Harvard University, Cambridge, MA 02138, USA}}
\affil[3]{\small{Institute for Mathematical and Computational Engineering, School of Engineering and Faculty of Mathematics, Pontificia Universidad Cat\'olica de Chile, Santiago, Chile}}
\date{\today}

\begin{document}
\maketitle

\begin{abstract}
We present a computational framework for efficient optimization-based ``inverse design'' of large-area ``metasurfaces'' (subwavelength-patterned surfaces) for applications such as multi-wavelength/multi-angle optimizations, and demultiplexers. To optimize surfaces that can be thousands of wavelengths in diameter, with thousands (or millions) of parameters, the key is a fast approximate solver for the scattered field.  We employ a ``locally periodic'' approximation in which the scattering problem is approximated by a composition of periodic scattering problems from each unit cell of the surface, and validate it against brute-force Maxwell solutions. This is an extension of ideas in previous metasurface designs, but with greatly increased flexibility, e.g. to automatically balance tradeoffs between multiple frequencies or to optimize a photonic device given only partial information about the desired field. Our approach even extends beyond the metasurface regime to non-subwavelength structures where additional diffracted orders must be included (but the period is not large enough to apply scalar diffraction theory).
\end{abstract}

%%%%%%%%%%%%%%%%%%%%%%%%%%  body  %%%%%%%%%%%%%%%%%%%%%%%%%%
\section{Introduction and motivation}
\label{sec:intro}

\begin{figure}[ht!]
\centering\includegraphics[width=13cm]{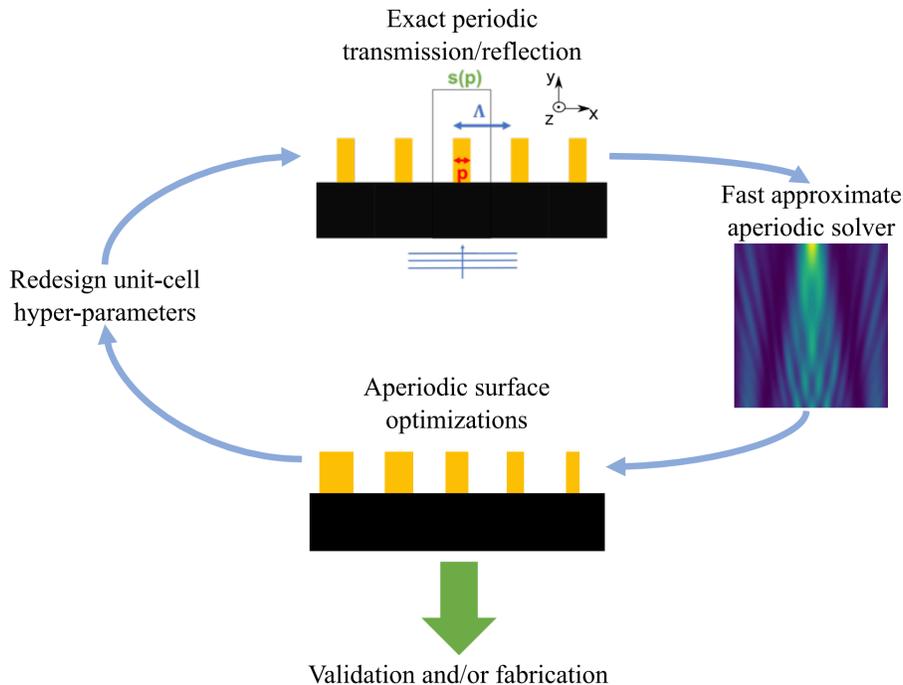}
\caption{Schematic of our design method: exact Maxwell scattering solutions for a set of periodic unit cells (top) are composed into an approximate solution for an arbitrary aperiodic composition (right), and this approximation is then used for large-scale optimization to determine the metasurface parameters to maximize a given objective (e.g. the focal intensity, bottom). The performance of the final design can then be fed back into adjusting the design of the unit cell (left).}
\label{fig:diag}
\end{figure}

In this paper, we present and validate a fast method for optimization-based ``inverse design'' of large (hundreds of wavelengths $\lambda$) aperiodic metasurfaces for wavefront shaping~\cite{lalanne_design_1999,yu_light_2011,kildishev_planar_2013,yu2013flat,jahani_all-dielectric_2016}, incorporating both scattered amplitude and phase for multiple incident $\lambda$ and angles. Previous methods either optimized the full Maxwell equations~\cite{lalau-keraly_adjoint_2013,piggott_inverse_2015,piggott_fabrication-constrained_2017,sell_periodic_2017, zhan2018inverse} (which is infeasible for large surfaces), were restricted to weakly coupled scatterers~\cite{matlack_designing_2018}, or started with a desired scattered phase and tried to design a corresponding metasurface unit cell~\cite{aieta_aberration-free_2012,arbabi_planar_2017, khorasaninejad_visible_2017, aieta_multiwavelength_2015, khorasaninejad_achromatic_2015, khorasaninejad_achromatic_2017, arbabi_controlling_2017,su2018advances,groever2018substrate, lin_topology_2017} 
(but if attainable unit cells fail to exactly match the desired $\lambda$-dependent phase there was no systematic way to choose the best compromise). In contrast, our approach starts with a family of manufacturable unit cells and directly optimizes an aperiodic composition for the desired field pattern by a fast approximate model, automatically finding the best compromise for the given constraints. Whereas phase-design methods typically assume that the desired scattered field is known everywhere~\cite{aieta_aberration-free_2012,arbabi_planar_2017, khorasaninejad_visible_2017, aieta_multiwavelength_2015, khorasaninejad_achromatic_2015, khorasaninejad_achromatic_2017, arbabi_controlling_2017,lin_topology_2017, su2018advances, groever2018substrate}, our approach allows one to specify the field objective only in regions of interest. 
%shall we speak about worst case optimization using phase-design?
 As outlined in Fig.~\ref{fig:diag}, given exact scattering calculations for small metasurface unit cells (\secref{subsec:library} and Fig.~\ref{fig:library_construction}), we build an approximate convolutional model of an arbitrary metasurface (\secref{subsec:equivalence}) that can then be optimized (\secref{sec:mid}) rapidly (seconds to find the optimum for a 200-$\lambda$ 2d aperiodic surface with hundreds of parameters) using two different objective functions.
We validate the optimized design with a brute-force Maxwell solver (\secref{subsec:wavefront}) and we find excellent quantitative agreement (Fig.~\ref{fig:greens_function}) even for rapidly varying aperiodic surfaces that challenge the assumptions of our model. We present example designs (\secref{sec:examples}) for a multi-wavelength optimization (Fig.~\ref{fig:rgb}), a wavelength demultiplexer (Fig.~\ref{fig:demultiplexer}), and an multi-angle optimization (Fig.~\ref{fig:angle}). Our approach is not limited to true ``metasurfaces'' whose features are small enough to mimic effective-impedance~\cite{achouri2015general,
epstein2014floquet,
epstein2014passive,
holloway2009discussion,
holloway2012overview,
kuester2003averaged,
pfeiffer2013metamaterial,
tcvetkova2018near,
tretyakov2015metasurfaces} surfaces: we show that it even works well for large-period microstructures that scatter multiple diffracted beams. Indeed, our method is easily extensible to incorporate multiple diffraction coefficients, multi-layer/multi-parameter unit cells, multiple polarizations, and other complications (\secref{sec:beyond} and Fig. \ref{fig:diffractive}).

\label{sec:model}

\section{Locally periodic approximation}
\label{sec:locallyperiodic}

The key to ``metasurface'' design is to be able to quickly calculate the transmitted/reflected field for a large-area structure, possibly thousands of wavelengths in diameter---too large to solve the full Maxwell equations without some simplifying assumption.   Similar to~\citeasnouns{aieta_aberration-free_2012,arbabi_planar_2017,
khorasaninejad_visible_2017, aieta_multiwavelength_2015, khorasaninejad_achromatic_2015, khorasaninejad_achromatic_2017, arbabi_controlling_2017, su2018advances,groever2018substrate}, the central approximation of our approach is to assume that the metasurface is \emph{locally periodic}: the scattering in any small region is almost the same as the scattering from a periodic surface. The use of periodic calculations to compute the specular reflection phase only, typically discarding amplitudes and additional diffracted orders, has sometimes been called a ``local phase approximation'' \cite{verslegers2010phase, cheng2017optimization}.  (Contrast this with the regime of scalar diffraction theory~\cite{born2013principles, o2004diffractive}, valid for period $\Lambda \gg$ wavelength $\lambda$, in which the surface is treated as locally \emph{uniform}, separately computing the transmission coefficient at each \emph{point} on the surface.) In a separate paper~\cite{carlos}, we develop the rigorous foundations and convergence rates of a related approximation, along with higher-order corrections, but here (similar to previous authors~\cite{aieta_aberration-free_2012,arbabi_planar_2017,
khorasaninejad_visible_2017, aieta_multiwavelength_2015, khorasaninejad_achromatic_2015, khorasaninejad_achromatic_2017, arbabi_controlling_2017, su2018advances}) we will simply perform brute-force Maxwell simulations at the end (\secref{sec:mid}) to validate our designs. (In fact, we will find in Fig.~\ref{fig:greens_function} that the locally periodic approximation gives excellent agreement with full simulations even for surfaces where the unit cell is rapidly varying in some regions.)
Unlike previous authors who calculated only the scattered phase and not the total scattered field~\cite{aieta_aberration-free_2012,arbabi_planar_2017,
khorasaninejad_visible_2017, aieta_multiwavelength_2015, khorasaninejad_achromatic_2015, khorasaninejad_achromatic_2017, arbabi_controlling_2017, su2018advances}, we employ both amplitude and phase information to formulate a complete approximate solver (scattered field for any given incident field) that can be used to optimize the metasurface for arbitrary ``objective'' functions of the field.  
Not only does this approximate solver enable very general optimization, it also allows us to evaluate the optimized metasurface for different (non-optimized) incident fields (e.g. the wavelength sensitivity computed in \secref{sec:examples} for the multi-wavelength lens).  As discussed in \secref{sec:beyond}, our approach is also easily extensible to a "non-metasurface" regime in which there are multiple diffracted beams from a large-period surface, as well as to computing near fields (via the scattering coefficients of the evanescent waves).  In this section, we explain in detail how the locally periodic approximation allows us to compute the total scattered/reflected field (at any point in space) for any incident wave.

To make it easier to understand our approach, it is helpful to consider a specific example of a two-dimensional ``metasurface'' unit cell, based on \citeasnoun{khorasaninejad_achromatic_2017}: TiO$_2$ pillars on top of a silicon dioxide substrate, as shown in Fig.~\ref{fig:library_construction}. The height of the pillar is fixed to 600 nm, the period $a$ is fixed to $235$~nm, and the pillar width varies: $p \in [50, a-50]$~nm (imposing a minimum feature size of 50~nm for practical fabrication). One could easily add more parameters and/or constraints, as discussed in \secref{sec:conclusion}.
Given this unit cell, an aperiodic metasurface is formed by taking a group of such unit cells with independent parameters and juxtaposing them next to each other. 

Our goal is to compute the scattered (transmitted or reflected) field for such an aperiodic surface, for any given incident wave (e.g. a planewave or gaussian beam), given \emph{only} the exact Maxwell solutions for scattering of planewaves by \emph{periodic} surfaces of the different pillar widths. In this paper, we consider only incident propagating (not evanescent) waves, but in another paper \cite{carlos} we show that a similar approach can be extended to evanescent fields as well.
The key ``locally periodic'' assumption is that the pillar width (the unit cell) changes sufficiently slowly from one pillar to the next.  (This assumption is rigorously quantified in~\cite{carlos}, is validated numerically in \secref{sec:examples}, and it turns out that we even obtain good accuracy when there are sudden changes in pillar width at a few locations.)
\edit{As mentioned above, this assumption is similar in spirit to other metasurface work~\cite{aieta_aberration-free_2012,arbabi_planar_2017, khorasaninejad_visible_2017, aieta_multiwavelength_2015, khorasaninejad_achromatic_2015, khorasaninejad_achromatic_2017, arbabi_controlling_2017, su2018advances}, where it was found to work well for a wide variety of metasurface designs; the main contribution of this paper is to couple the locally periodic approximation to general optimization tools and near-to-far-field transformations. Of course, this approximation can break down for devices requiring extremely rapid surface variations such as diffraction to nearly glancing angles~\cite{kim2018design,carlos}, although it can be generalized by including a next-order correction~\cite{carlos}, but this is not a problem for the moderate-NA lens-like applications considered in this paper.}

\subsection{Periodic sub-problems}
\label{subsec:library}
%add local field
\begin{figure}[ht!]
\centering\includegraphics[width=13cm]{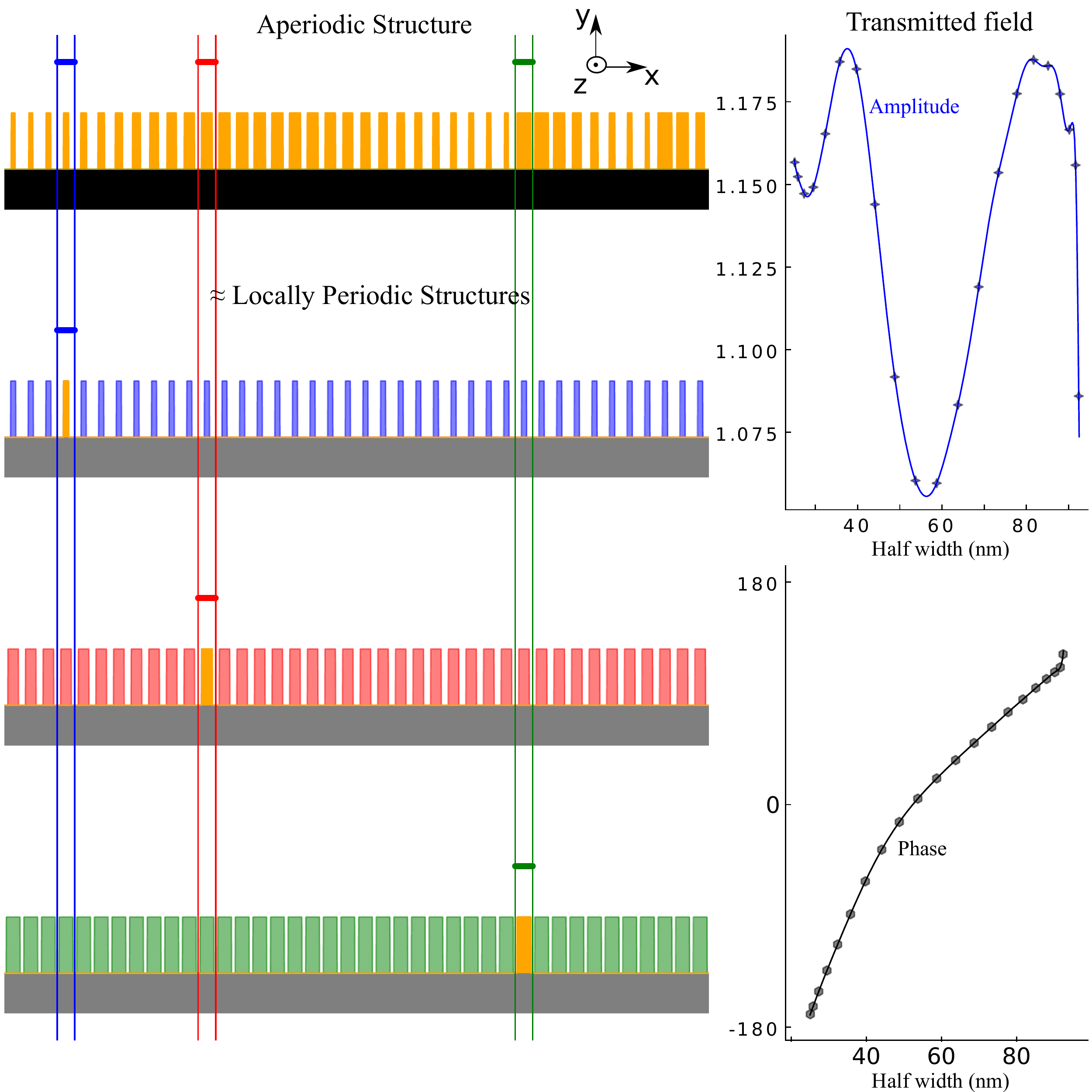}
\caption{Left: an arbitrary aperiodic metasurface (top) is approximated by solving a set of periodic scattering problems (bottom), one for each unit cell, to obtain the scattered field just above the surface (horizontal line segments).  Right: 0th diffracted-order amplitude (top) and phase (bottom) of periodic subproblems as a function of the pillar width. This is precomputed for several widths (markers) and interpolated as needed.}
\label{fig:library_construction}
\end{figure}

%paragraph 1 (about left part of the figure)
In Fig.~\ref{fig:library_construction}(left) is shown the fundamental assumption of our approach. For each unit cell of the aperiodic structure, we approximate the field in a plane/line just above the unit cell by the solution for the equivalent periodic structure. Three examples are highlighted corresponding to three different parameters of the unit cell.
When the period of the unit cell is subwavelength, the zeroth diffractive order is the only propagating wave~\cite{Joannopoulos:2008:PCM:1628775}. 
Therefore, if we are interested only in the far field, we can make an additional approximation: we replace the scattered field by its zeroth Fourier component which is simply the average of the field on the plane just above the pillar.
Given this approximate field just above the surface, in \secref{subsec:equivalence} we construct an approximate field everywhere above the surface. 
In \secref{sec:beyond}, we go beyond the zeroth-order (specular) approximation by including additional diffractive orders.

%paragraph 2
We will consider periodic structures with hundreds of different pillar widths (with a fixed period), but we would like to avoid having to do hundreds of unit-cell calculations. 
We can take advantage of the fact that the scattered fields are smooth functions of the pillar width \cite{costabel2012shape, costabel2012shape2} by solving the scattering function for a few widths and then interpolating to any other widths.

Given a smooth function $f(p)$ of some parameter $p_1 \le p \le p_2$, Chebyshev methods evaluate $f(p)$ at a few special points $p$ and construct a polynomial approximation $\tilde{f}(p)$ that can be used to rapidly evaluate the $f(p)$ with exponentially good accuracy. This can be extended to multiple parameters using products of Chebyshev polynomials~\cite{boyd2001chebyshev} or by more sophisticated methods such as sparse grids~\cite{Gerstner1998}.  In this way, we only need to solve the unit-cell Maxwell problem a few times to obtain our polynomial approximant $\tilde{f}(p)$, which is then evaluated, along with its derivative, many times during optimization. In particular Fig.~\ref{fig:library_construction}(right) shows the real part, the imaginary part, and the phase of the zeroth Fourier coefficient of the transmitted field versus the pillar width. 
We evaluate this coefficient for 21 different widths (at Chebyshev points~\cite{boyd2001chebyshev}), and can then interpolate to high accuracy using Chebyshev polynomials~\cite{boyd2001chebyshev}.  Multiple parameters per unit cell could be interpolated using a product of Chebyshev polynomials~\cite{boyd2001chebyshev} or, for more than 3--4 parameters, sparse-grid interpolation~\cite{Gerstner1998}. Here, we use the finite-difference frequency-domain (FDFD) method~\cite{CHAMPAGNE2001830, maxwellfdfd-webpage} with perfectly matched layer (PML) absorbing boundaries~\cite{SHIN20123406, Mittra1995}, but any other computational method for periodic Maxwell problems would work as well.

\subsection{Green's functions and the equivalence principle}
\label{subsec:equivalence}

%paragraph 1 current
Once the fields are known in a plane above the metasurface, we can obtain the fields \emph{everywhere} above the metasurface using the \emph{principle of equivalence}~\cite{harrington, oskooi2013electromagnetic}, \edit{also known as a near-to-far-field transformation~\cite{taflove2005computational})}: the fields in the \edit{$y=y_0$} plane can be treated as equivalent current sources that generate the fields everywhere else.  These equivalent electric ($\vec{J}$) and magnetic ($\vec{K}$) current densities are defined by~\cite{oskooi2013electromagnetic}:
\begin{eqnarray}
\begin{bmatrix}
\vec{J}\\
\vec{K}
\end{bmatrix} = \delta(y-y_0) \begin{bmatrix}
\hat{\vec{n}} \times \vec{H}\\
-\hat{\vec{n}} \times \vec{E}
\end{bmatrix}
\label{eq:1}
\end{eqnarray}
where $\hat{n} = \hat{y}$ is the surface unit-normal vector and the delta function implies that these are surface currents on the plane $y=y_0$.

%paragraph 2 simplification
A further simplification is possible if we only care to compute the fields \emph{above} the surface.  The currents~\eqref{eq:1} produce the desired scattered fields above the plane and zero fields below the plane~\cite{harrington, oskooi2013electromagnetic}, and this means that the same fields above are produced if we add or subtract the mirror-image currents (which produce fields below and zero above). Subtracting the mirror-image sources, however, cancels the $\vec{J}$ term and leaves only the $\vec{K}$ current (a pseudovector under mirror flips~\cite{jackson_classical_1999}).
This allows us to use only $\hat{\vec{n}} \times \vec{E}$ sources arising from the electric field computed by the locally periodic approximation in section \ref{subsec:library}.   As explained in section~\ref{subsec:library}, we can further approximate the $\vec{E}$ field by its average in each unit-cell calculation for subwavelength periods, since this gives the far-field diffracted order.

%paragraph 3 integrating against the Maxwell Green's function  [ref]
Given these equivalent currents, or their approximation by far-field locally periodic calculations, the electric (or magnetic) fields at any point $\vec{x}$ above the surface can be computed by integrating along with the Maxwell Green's function (the field at $\vec{x}$ from a source at $\vec{x}'$)~\cite{harrington}. 
For our two-dimensional model problem ($xy$ plane) with the $E_z$ polarization, where we only have a current $K_x(\vec{x}') = -E_z(\vec{x}') \delta(y-y_0)$ and we let $G(\vec{x}, \vec{x}')$  denote the relevant component of the Green's tensor [$E_z(\vec{x})$ from $K_x(\vec{x}')$], this integral takes the form
\begin{equation} 
E_z(\mathbf{x}) = - \int_\mathrm{surface} G(\mathbf{x}, \mathbf{x}') E_z(\vec{x}') \, d\mathbf{x}'
\label{eq:2}
\end{equation}
where $G$ is a Hankel function~\cite{watanabe_integral_2014} $G(\mathbf{x}, \mathbf{x}^\prime) = -\frac{ik}{4}H_1^{(1)}(kr) \hat{\vec{n}} \cdot \frac{\vec{r}}{r}$, where $k  = \frac{2\pi}{\lambda}$, $\vec{r} = \vec{x}-\vec{x}'$, and $r=|\vec{r}|$.
For a finite metasurface with an infinite silica substrate, we use a standard ``windowing'' method to truncate this integral accurately to a finite region~\cite{bruno_windowed_2016, bruno_windowed_2017, bruno_windowed_2017-1}.

This equivalent-currents formulation is exact if the true aperiodic $E_z$ field is used for the $K_x$ source term, and in section \ref{sec:mid} we find that it has excellent accuracy with the locally periodic approximation for typical metasurface designs.  (A related approximation is made in scalar diffraction theory, where the locally \emph{uniform} approximate scattered fields can be thought of as sources producing fields everywhere else~\cite{born2013principles, o2004diffractive}, further approximated in the far field by e.g. the Fraunhofer diffraction theory~\cite{born2013principles}.)

\section{Metasurface inverse design methods}
\label{sec:mid}

The previous section gives us a fast way to solve the forward problem for the scattered field above a given metasurface.
In this section, we see how we use it to solve the inverse problem, i.e. find the parameters of a metasurface to produce a desired scattered field.
We solve it as an optimization problem: we minimize or maximize an \emph{objective} function of the unit cell parameters subject to some constraints.
\edit{Given an efficient way to compute any objective function and its gradient, there are a wide-variety of well-known optimization methods that can be applied;
we use the ``CCSA-MMA'' algorithm~\cite{svanberg_class} via a free-software implementation \cite{johnson2014nlopt}.}
To avoid getting trapped in poor local optima, we use a technique called successive refinement~\cite{MutapcicBo09, chan2000multilevel, chun1992fast}: 
We successively double the number of degrees of freedom, using the optimized coarser structures as starting points for optimization of the finer structures.
\edit{(The result was not very sensitive to the starting parameter guess; we simply started each parameter in the middle of its allowed range.)}
What objective function should we optimize?
In \secref{subsec:wavefront} we consider an objective function similar to previous work~\cite{aieta_multiwavelength_2015,lin_topology_2017}, which matches the field just above the metasurface with the desired field.
In \secref{subsec:intensity} we optimize more general functions of the scattered field, e.g. the intensity at a single focal point, which is more flexible when only partial information is known about the desired field.
In both cases, in this section we will design a simple lens structure that we will validate using brute force simulations. In \secref{sec:examples}, we will consider more difficult design problems.
In \secref{subsec:minmax} we generalize our approach to multiple frequencies and angles of incidence via a maxmin formulation.

\subsection{Optimizing the wavefront}
\label{subsec:wavefront}
\begin{figure}[ht!]
\centering\includegraphics[width=10cm]{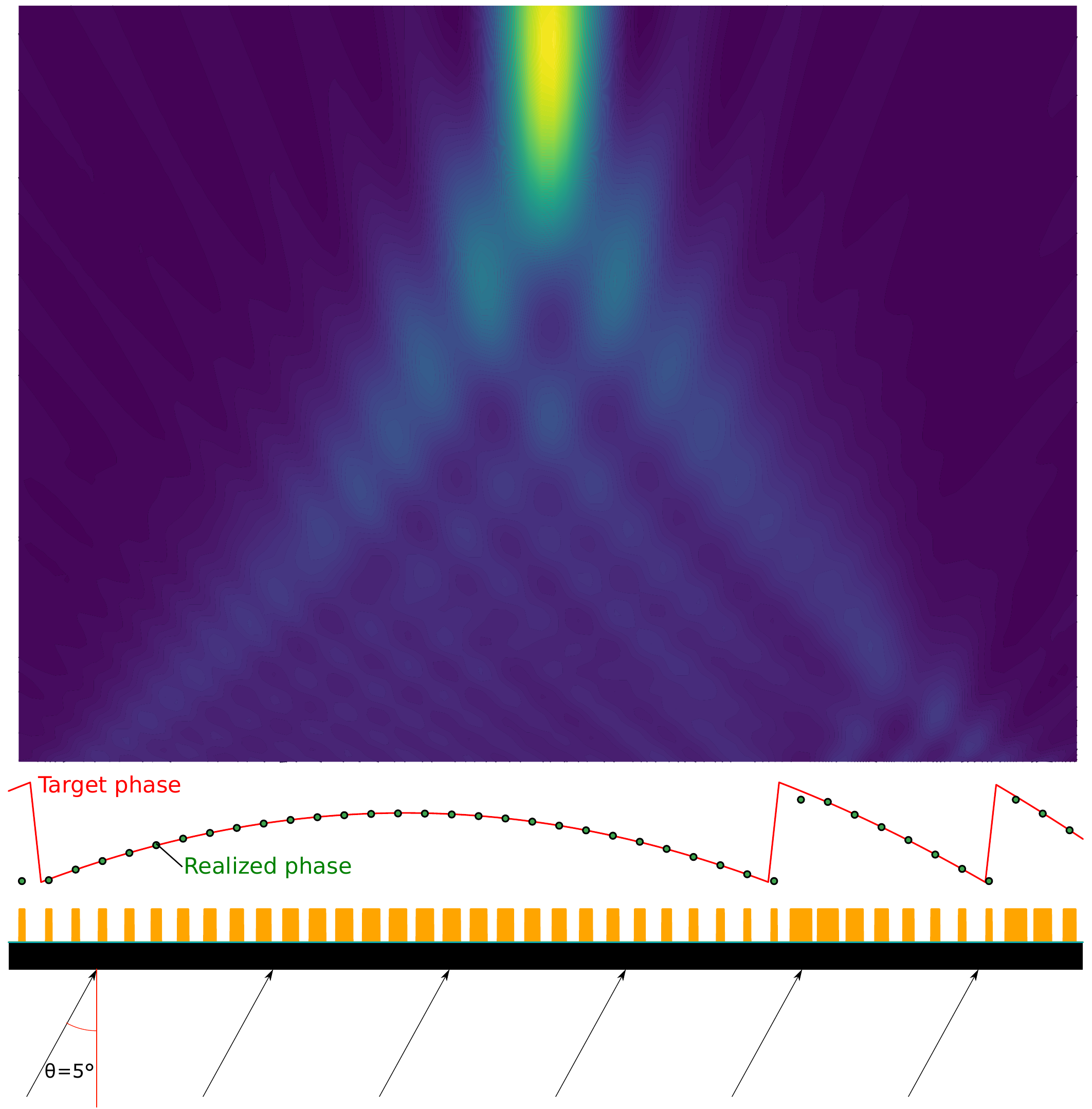}
\caption{Bottom: geometry of a metasurface designed for a 5-degree incident plane wave of wavelength 532~nm and focal length 14.7~$\mu$m (numerical aperture of 0.3) using the wavefront method. This design produces a field with the needed phase (middle). Top: $|E_z|^2$ intensity plot shows focusing to the target focal spot.}
\label{fig:wavefront_method}
\end{figure}

\begin{figure}[ht!]
\centering\includegraphics[width=12cm]{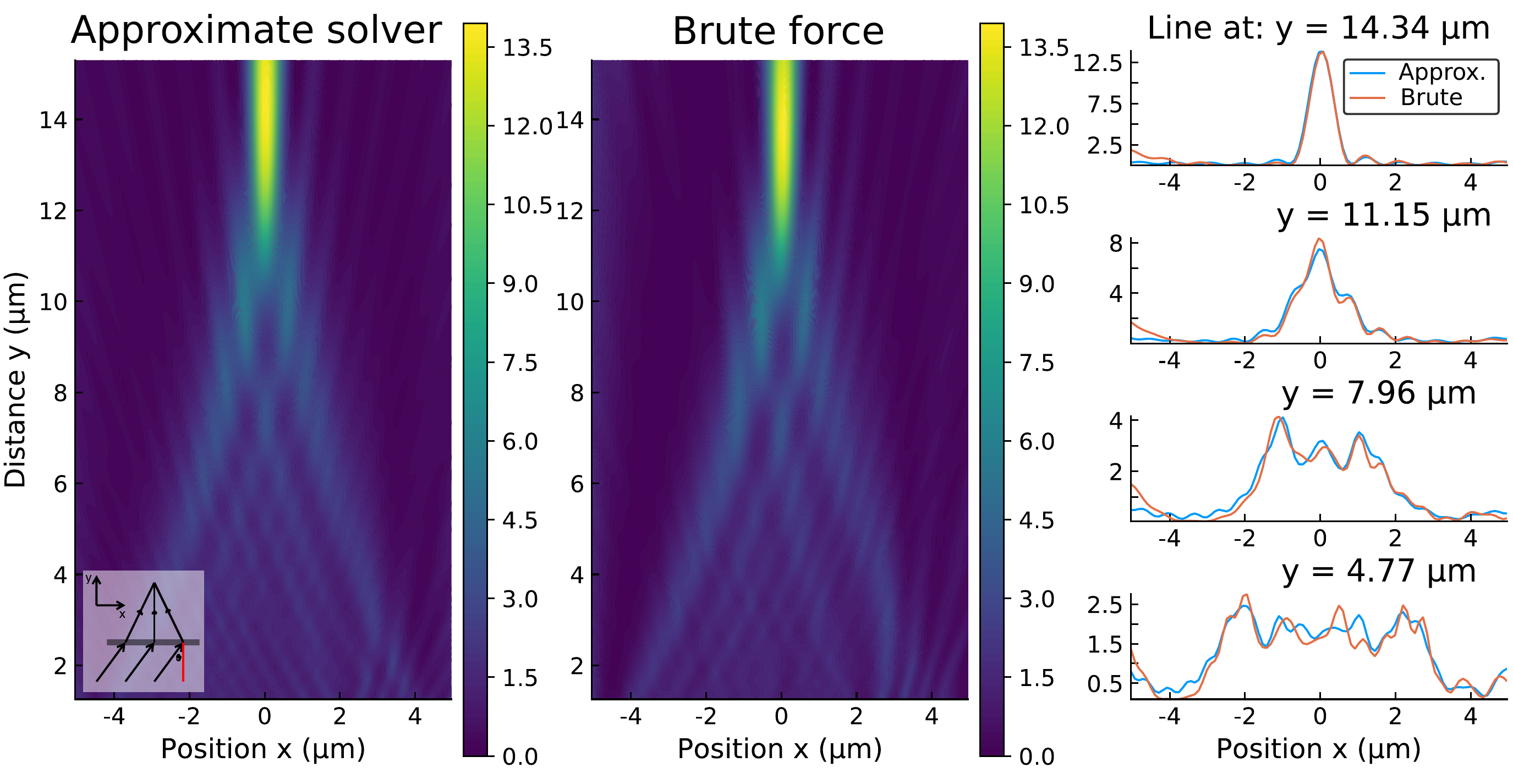}
\caption{Left: $|E_z|^2$ intensity plot of the scattered field of a metalens using our locally periodic approximate solver showing good agreement with a brute force calculation (middle). Right: the field sections computed by the two solvers show perfect agreement close to the focal lines (top). This agreement starts to deteriorate the closer the section is to the metasurface (bottom).}
\label{fig:greens_function}
\end{figure}

%paragraph 1 the optimization problem
When the exact desired field is known everywhere above the metasurface, as in lens design~\cite{
aieta_aberration-free_2012,
arbabi_planar_2017,
khorasaninejad_visible_2017, 
aieta_multiwavelength_2015, 
khorasaninejad_achromatic_2015, 
khorasaninejad_achromatic_2017, 
arbabi_controlling_2017, 
groever2018substrate,
su2018advances, lin_topology_2017} and other wavefront shaping problems, by the equivalence principle~\cite{harrington} it is sufficient to produce this field on a plane just above the surface. Since the approximate scattered fields $s(p)$ just above the surface (the $E_z$ produced for a given metasurface parameter $p$) are given by the locally periodic approximation in \secref{subsec:library}, we can directly minimize the difference between this $s$ and the desired field $a(x)e^{i\phi(x)}$:
\begin{equation}
\min_{s_0, \phi_0, p} \int |s(p(x))- s_0 a(x) e^{i \phi(x) + i\phi_0}|^2  dx ,
\label{eq:wavefront}
\end{equation}
where $s_0$ and $\phi_0$ are an unknown overall amplitude/phase and $p(x)$ describes the metasurface parameters along the surface.  This approach eliminates the need for any Green's function integral (\secref{subsec:equivalence}) to obtain the field elsewhere. 

For a lens application, typically $a(x) = 1$ and all of the information is in the desired phase $\phi(x)$~\cite{aieta_multiwavelength_2015}. A closely related approach was used for metalens design in several previous works ~\cite{aieta_aberration-free_2012,arbabi_planar_2017,
khorasaninejad_visible_2017, aieta_multiwavelength_2015, khorasaninejad_achromatic_2015, khorasaninejad_achromatic_2017, arbabi_controlling_2017, su2018advances, lin_topology_2017}. There, since both $a(x)$ and the locally periodic far-field $|E_z|$ were approximately constant, the amplitude was ignored and they simply attempted to match the desired phase. If this phase can be matched exactly in a given unit cell by tuning its parameter $p(x)$ (e.g. pillar width), then no explicit optimization formulation is needed~\cite{aieta_aberration-free_2012}, but an optimization-based approach is more flexible at balancing tradeoffs in cases where the desired $ae^{i\phi}$ cannot be exactly obtained, especially in multi-frequency problems (\secref{subsec:rgb}).  A phase-based optimization approach was directly employed in \citeasnoun{lin_topology_2017} for topology optimization of a small area (no locally periodic approximation).

%paragraph 2 our solution
For example, in Fig.~\ref{fig:wavefront_method} we minimize equation~\eqref{eq:wavefront} for a single-frequency $\lambda = 532\,\mathrm{nm}$ lens design problem: we focus an incident planewave at a 5-degree angle on a focal point $14.7\,\mathrm{\mu m}$ from the surface, using the target phase $\phi(x)$ from \citeasnoun{aieta_multiwavelength_2015}.  We optimize over piecewise-constant parameters, effectively one parameter per unit cell, with a standard optimization algorithm~\cite{svanberg_class} utilizing analytically computed gradients of the objective function with respect to the parameters.  Starting the optimization from a constant-$p$ initial guess was sufficient to obtain a local minimum with excellent performance shown in Fig.~\ref{fig:wavefront_method}. (This 40 unit-cell optimization required $<100$~ms on a laptop.)   At the top is the $|E_z|^2$ intensity plot computed with by our approximate solver (\secref{subsec:equivalence}). Below this is shown the $96\%$ match between the desired and obtained fields (from the locally periodic approximation) just above the metasurface.  At the bottom is shown the optimized metasurface geometry, which is mostly slowly varying but has sudden jumps in the pillar widths when the desired phase passes through $2\pi$.

In Fig.~\ref{fig:greens_function}, the locally periodic approximate solver (left) is compared to a brute-force surface-integral equation (SIE) Maxwell solver~\cite{bruno_windowed_2016, bruno_windowed_2017, bruno_windowed_2017-1} for this optimized solution, showing good quantitative agreement.  More precisely, at right we compare the computed intensities $|E_z(x,y)|^2$ for several separations $y$ from the surface.   On the focal line, the mean squared difference between the solutions divided by the mean squared intensity is only $0.3\%$, validating our locally periodic approximation.    The errors increase as one approaches the surface because of effects that decay with distance---scattered waves (intensity $\sim 1/y$) from sudden jumps in the pillar width (which violate the locally periodic approximation) along with evanescent fields that we neglected in our far-field approximation---combined with the fact that small errors are more apparent in low-intensity regions far from the focal point.

\subsection{Optimizing arbitrary functions of the field}
\label{subsec:intensity}

Alternatively, since \secref{subsec:equivalence} allows us to compute the approximate field anywhere above a metasurface, we can optimize \emph{any} function of this field. This is especially useful if the desired field is only partially known: perhaps one cares about the field in some regions but not others, or is interested in amplitude but not phase.  In particular, here we approach the lens-design problem by directly maximizing the intensity $|E_z(\vec{x})|^2$ at a single focal point $|\vec{x}|$, which can be rapidly computed by a single integral \eqref{eq:2} of the locally periodic surface fields.  As in the previous sections, we used standard optimization techniques~\cite{svanberg_class} with an analytically computed gradient (essentially via an adjoint method~\cite{strang2007computational}), and the optimized structure for 40 unit cells was found in $< 1$~s on a laptop (whereas our brute-force solver was about $10^5$ times slower).   A comparison of the two methods when the period is not sub-wavelength appears in \secref{sec:beyond}.

\subsection{Max--min multi-objective optimization}
\label{subsec:minmax}

Many design problems involve a combination of multiple objectives: maximizing performance at different wavelengths, angles, and/or focal spots, for example.  One common way to do this is a \emph{max--min} formulation: we optimize the worst objective:
\begin{equation*}
\begin{aligned}
& \max_\mathrm{parameters} 
& &  \left[ \min_{\lambda\in\mathrm{wavelengths}} \mathrm{objective}(\mathrm{parameters}, \lambda) \right] .
\end{aligned}
\end{equation*}
(For example, in \secref{subsec:rgb}, the ``objective'' function for an RGB lens is the intensity at the focal spot, and max--min optimization means that we try to maximize the \emph{lowest} intensity across the three design wavelengths.)
Although the expression $[\cdots]$ being maximized is no longer differentiable, which would make the most efficient high-dimensional optimization methods inapplicable, it can be transformed into an equivalent differentiable problem~\cite{Boyd:2004:CO:993483}
\begin{equation*}
\begin{aligned}
& \max_{t,\mathrm{parameters}} 
& & t \\
& \text{subject to}
& & t \leq \mathrm{objective}(\mathrm{parameters}, \lambda) \mbox{ for } \lambda\in\mathrm{wavelengths}.
\end{aligned}
\end{equation*}
where $t \in \mathbb{R}$ is a new ``dummy'' optimization parameter.  Assuming that the original objective function is differentiable, we can now use a standard nonlinear constrained-optimization algorithm~\cite{svanberg_class}.  \edit{In particular, the CCSA-MMA algorithm~\cite{svanberg_class} only requires us to supply the functions $t$ and $t - \mathrm{objective}(\mathrm{parameters}, \lambda)$ and their gradients (with respect to $t$ and the parameters) in order to solve the local-optimization problem. Efficient gradient formulas for our cost functions from \secref{subsec:wavefront} and \secref{subsec:intensity} are given in the Appendix.}

We will show examples of such optimization problems in \secref{sec:examples}, where we will use max--min to optimize for multiple frequencies (Fig.~\ref{fig:rgb} and Fig.~\ref{fig:demultiplexer}) or angles (Fig.~\ref{fig:angle}).

\section{Applications: RGB lens, demultiplexer, and angle-insensitive lens}
\label{sec:examples}

In this section, we show some larger and more interesting design problems that can be solved by our methods from the previous section.  We still use the same TiO$_2$ pillar unit cells as in \secref{sec:locallyperiodic}, but now we consider metasurfaces consisting of 1000 unit cells, combining multiple frequencies and/or angles, and we could solve the resulting optimization problems in a few minutes on a laptop. In particular, we consider three applications: a lens which has the \emph{same} focal spot for RGB (red, green, blue) wavelengths, a demultiplexer that focuses RGB wavelengths at three \emph{different} focal spots, and a lens that focuses four incident \emph{angles} at the same wavelength to the same focal spot. We will also show that our methods are suitable for sensitivity analyses with respect to wavelengths or angles, by evaluating designs at non-optimized inputs using our fast (locally periodic) solver.

\subsection{Max--min RGB (red, green, blue) focusing}
\label{subsec:rgb}

%problem
Here, we use the max--min method of \secref{subsec:minmax} to focus normally incident plane waves of three different wavelengths---480~nm (blue), 530~nm (green), and 650~nm (red)---on a \emph{single} focal spot, by maximizing the minimum (worst) intensity at that spot for all three wavelengths. The diameter of the lens is 235~microns (1000 unit cells), and the focal length is 350.6~microns, which corresponds to a numerical aperture of 0.3.

\begin{figure}[ht!]
\centering\includegraphics[width=13cm]{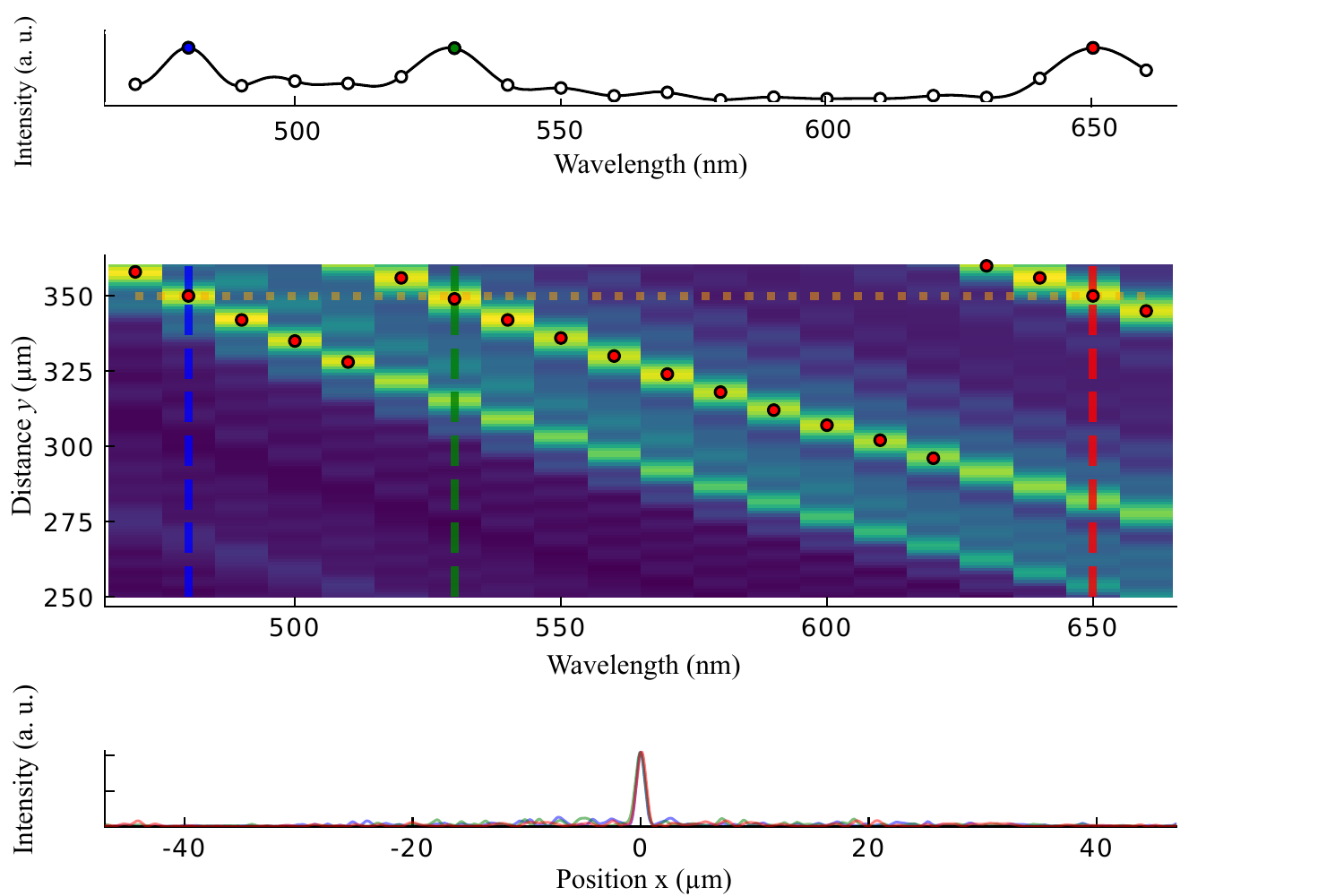}
\caption{Bottom: the focal line of the scattered field for the three target wavelengths (blue, green, and red) show a clear focusing on the target focal axis. Middle: sensitivity plot for the focal length with respect to the wavelength show chromatic aberration, and each wavelength objective creates a ``spurious focus'' (local maximum along along the focal axis) on the focal axis at other wavelengths. The red spots represent the foci for each wavelength, and we clearly see chromatic aberration. Top: intensity at the target spot vs.~wavelength.}
\label{fig:rgb}
\end{figure}

%result $|E_z|^2$
At the bottom of Fig.~\ref{fig:rgb} is shown the intensity on the focal line for all three wavelengths, demonstrating nearly diffraction-limited focusing (RGB half-maximum widths of 975, 997, and 850~nm, respectively).  In Fig.~\ref{fig:rgb}(middle), we evaluate our optimized design along the focal axis (a fixed $x=0$) versus distance $y$ from the surface and versus wavelength across the visible spectrum, in order to show the wavelength sensitivity of our RGB design. This plot reveals that the optimized design is actually producing three different focal spots (local intensity maxima) on the focal axis for every wavelength, and at each of the RGB wavelengths a different spot is brought to the 350.6~$\mu$m target.   At this target focal spot, the intensity $|E_z|^2$ is plotted versus wavelength in Fig.~\ref{fig:rgb}(top), showing the narrowband nature of the RGB focus.   The ability of our approximate solver to rapidly evaluate the performance of the design with many different (non-optimized) inputs ($<100$~ms each) is a powerful tool for characterizing and understanding the metasurface.

\subsection{Demultiplexer}
\label{subsec:demultiplexer}

%problem
Here, we design a \emph{demultiplexer} that focuses normally incident plane waves of three different wavelengths (RGB again) at three \emph{different} points, which are sixty microns laterally ($x$) apart from each other on the same focal plane (again 350.6~$\mu$m from the surface, a numerical aperture of 0.3). As above, we use the max--min formulation from \secref{subsec:minmax} to maximize the worst case intensity at the focal spots.

\begin{figure}[ht]
\centering\includegraphics[width=11	cm]{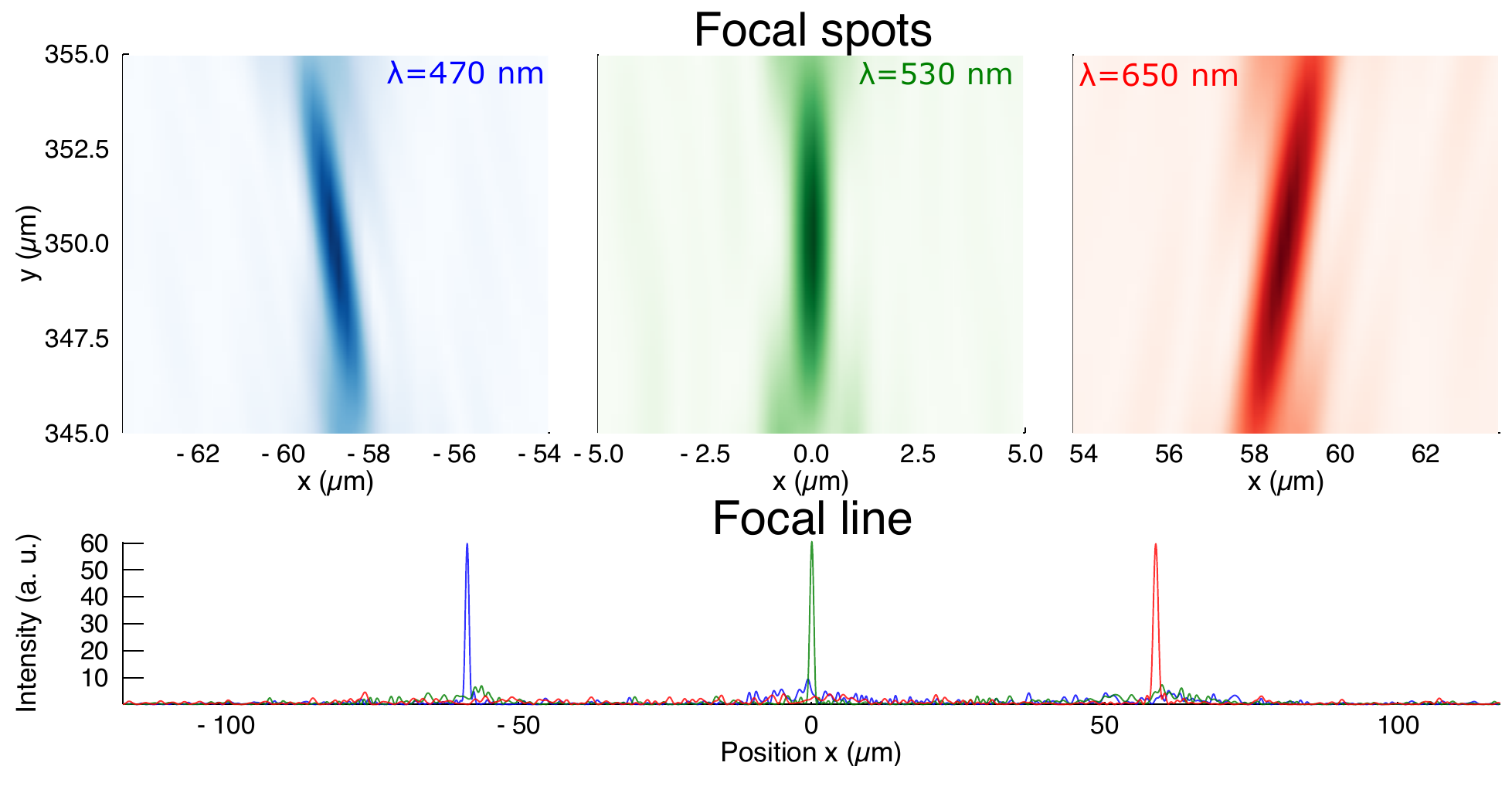}
\caption{Bottom: focal lines for the three target wavelengths (blue, green and red) focus on points sixty microns apart. Top: the field produced by our design focuses on the desired foci, the high-intensity regions for blue (left) and red (right) are tilted because their foci are off-axis.}
\label{fig:demultiplexer}
\end{figure}

%solution
In Fig.~\ref{fig:demultiplexer}(top), we show the field intensities in the vicinity of the three focal spots for the RGB wavelengths, and in Fig.~\ref{fig:demultiplexer}(bottom) we plot the corresponding intensities along the focal line $y=350.6\,\mu$m.  The focal spots for the two side focal points are tilted outward from the focal axis, which makes sense because they required off-axis focusing relative to the center of the metasurface.  As in \secref{subsec:rgb}, we attain nearly diffraction-limited RGB foci half widths of 825, 785, and 795~nm, respectively.

\subsection{Max--min multi-angle focus}

Our last application is a metasurface focusing incident plane waves coming at four different angles of incidence (normal $0^\circ$, $3^\circ$, $6^\circ$, and $9^\circ$) at the same focal point for the wavelength 532~nm, inspired by earlier topology-optimization work~\cite{lin_topology_2017}. As in the previous sections, we target a focal length of 350.6~$\mu$m (numerical aperture 0.3), and use the max--min formulation of \secref{subsec:minmax} to maximize the worst-case focal-point intensity.

\begin{figure}[ht!]
\centering\includegraphics[width=12cm]{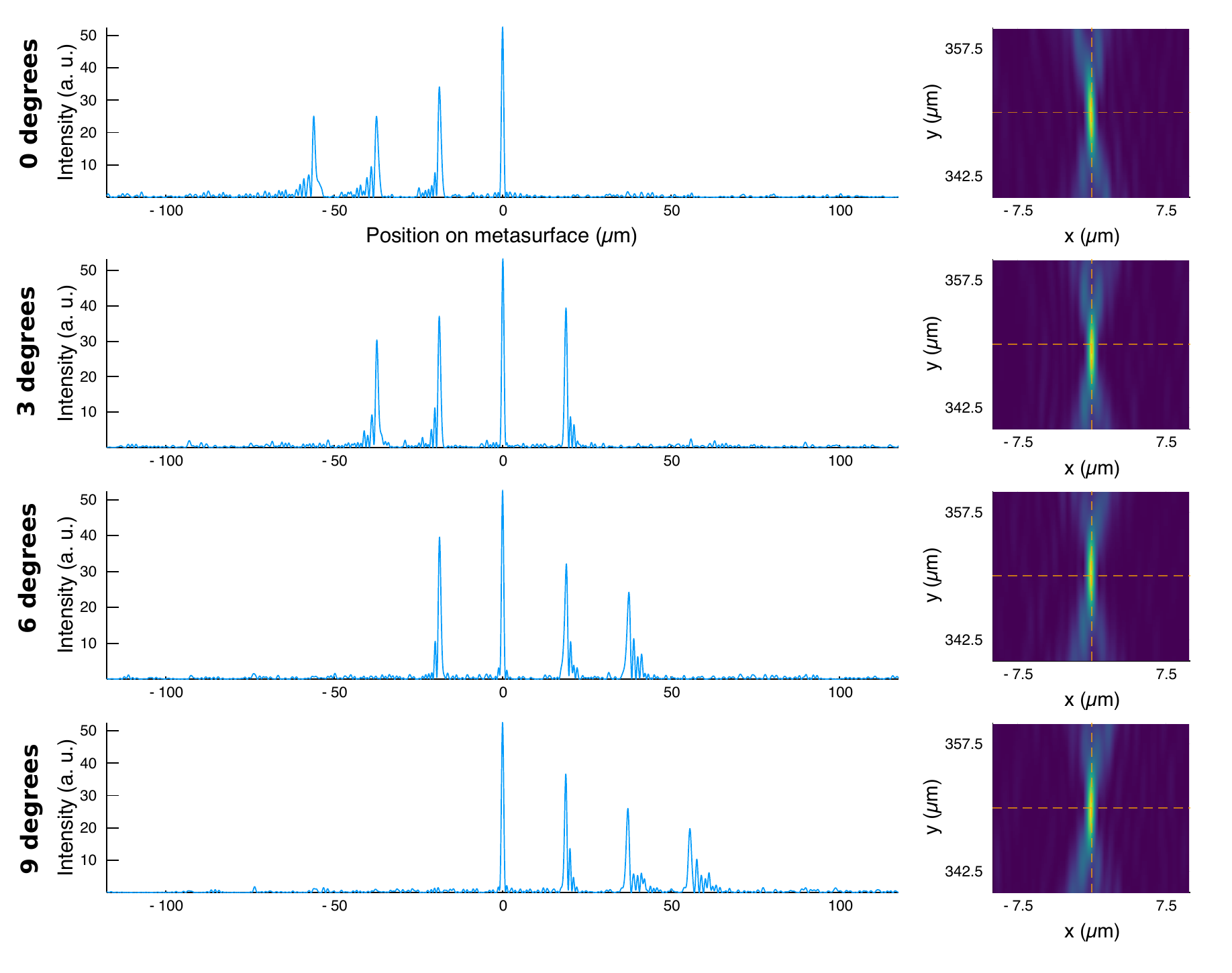}
\caption{Left: the focal lines for 0-degrees, 3-degrees, 6-degrees, and 9-degrees angles of incidence show four foci with the maximum intensity at the target focal spot (at $x=0$), the other three peaks correspond to the other three target angles. Right: corresponding produced field around the foci, the focal spot becomes more tilted as the angle of incidence increases. }
\label{fig:angle}
\end{figure}

Fig.~\ref{fig:angle}(right) shows the field intensities in the vicinity of the target focal spot for the four angles, exhibiting an unsurprising ``tilt'' proportional to the angle of incidence.   As in the previous sections, the spots are nearly diffraction limited (half widths of 787, 787, 807, and 724~nm).  Fig.~\ref{fig:angle}(left) shows the corresponding intensities on the focal plane $y=350.6\,\mu$m versus~$x$.   This plot shows that, in addition to a peak at the target point $x=0$, the metasurface produces three auxiliary side peaks.  (Preliminary work indicates that, similar to \citeasnoun{lin_topology_2017}, these auxiliary peaks can be mostly eliminated by redesigning the unit cell via additional parameters; we will address this in a future manuscript.) That is, much like in Fig.~\ref{fig:rgb}, the metasurface is creating four focal spots, such that at each angle of incidence a different focal spot is brought to the $x=0$ target point. The complex surface design and resulting transmitted field here would be very difficult to reproduce without large-scale optimization.

\section{Beyond subwavelength periods}
\label{sec:beyond}

%problem
The term ``metasurface'' should strictly apply only to deeply subwavelength structures that can be accurately described by an effective surface impedance/admittance or similar~\cite{achouri2015general,
epstein2014floquet,
epstein2014passive,
holloway2009discussion,
holloway2012overview,
kuester2003averaged,
pfeiffer2013metamaterial,
tcvetkova2018near,
tretyakov2015metasurfaces}, and most previous work operated in a subwavelength regime~\cite{
aieta_aberration-free_2012,
arbabi_planar_2017,
khorasaninejad_visible_2017, 
aieta_multiwavelength_2015, 
khorasaninejad_achromatic_2015, 
khorasaninejad_achromatic_2017, 
arbabi_controlling_2017, 
su2018advances}. Conversely, when the period is larger than the wavelength, additional diffracted waves appear in the far field~\cite{Joannopoulos:2008:PCM:1628775} that cannot be described by a uniform effective medium or by a single Fourier coefficient. Nevertheless, if the unit cells are mostly slowly varying it should still be valid to describe the surface by a locally periodic approximation (analogous to the adiabatic theorem for propagation through nearly periodic media~\cite{JohnsonBi02}) to approximate the field just above the surface and hence the field everywhere as in \secref{sec:locallyperiodic}.   When we solve the local periodic problems in non-subwavelength structures we can no longer retain only the 0th order Fourier coefficient, but instead we must retain either the full $E_z$ field on the surface or, for far-field calculations, the Fourier coefficients corresponding to all of the non-evanescent diffracted orders~\cite{Joannopoulos:2008:PCM:1628775}.

\begin{figure}[ht!]
\centering\includegraphics[width=10cm]{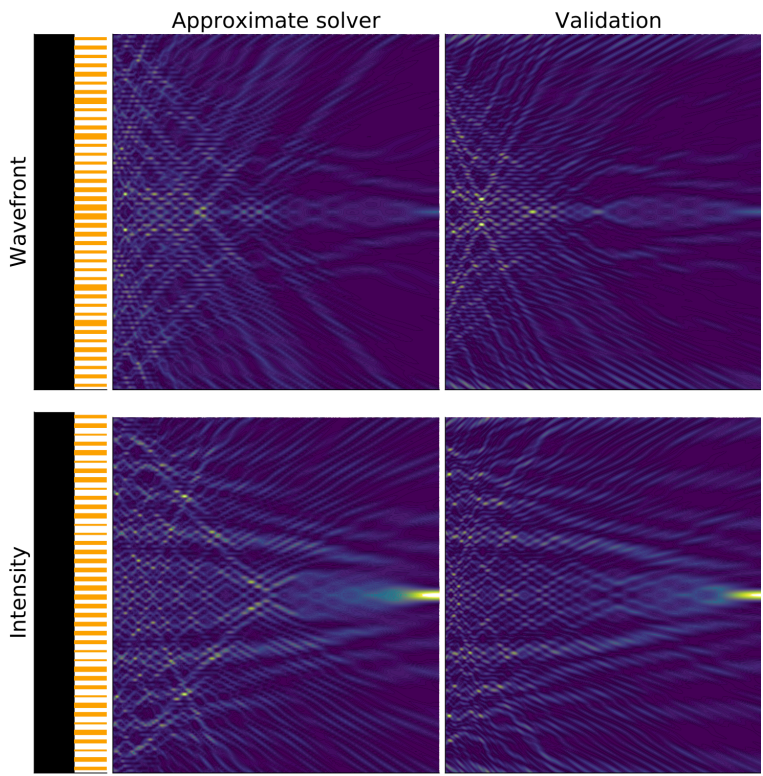}
\caption{Bottom: the geometry (left) from intensity optimization shows big variations in the width of the pillar, and produce good focusing when simulated with a brute force simulation (right), or our locally periodic solver (middle) which includes the diffractive orders $\pm 1$. Top: the geometry (left) from wavefront optimization shows poor focusing both using our locally periodic solver (middle) or a brute force calculation (right). All the intensity plots have the same color scale.}
\label{fig:diffractive}
\end{figure}

In Fig.~\ref{fig:diffractive} we show a single-wavelength ($\lambda=532$~nm) lens design for a period of $800\,\mathrm{nm} > \lambda$, so that even a periodic surface produces two additional diffracted orders $\pm 1$ in addition to the 0th-order ``specular'' transmission.   (Other than the period, the structure is the same TiO$_2$ pillar geometry considered in the previous sections, we use normal incidence, and design for a focal length of 48.6~$\mu$m with 40 unit cells similar to \secref{sec:locallyperiodic}.)  We considered both the wavefront and the intensity optimization approaches, and validated against a brute-force Maxwell solution as in \secref{sec:mid}.   Since the additional diffracted orders propagate at oblique angles, they have little influence on the focal intensity if the lens is designed to focus the 0th-order (specular) transmitted wave sufficiently far from the surface, so we carry out the inverse design using only the 0th-order term in the approximate model.

The results in Fig.~\ref{fig:diffractive} show that the intensity method still produces an excellent (near diffraction-limited) focal spot with high intensity that agrees well with the brute-force validation, whereas the wavefront optimization produces a much weaker focus that agrees poorly with the validation.  In both cases, the brute-force calculation and the approximate solver (which includes also the diffractive orders $\pm 1$) clearly show the additional diffracted orders scattering to oblique angles that have low amplitude at the focal spot. 
One major problem with the wavefront approach in this geometry is that varying the pillar width in this case changes the amplitude from 1 to 0.2, very different from the constant amplitude $\approx1$ in the subwavelength case. The best phase match corresponds to a weak efficiency, whereas the intensity method can compensate by utilizing both amplitude and phase variations.  The resulting lens designs shown in Fig.~\ref{fig:diffractive}(left) correspondingly have an average amplitude twice bigger for the intensity approach than for the wavefront approach (0.8 vs 0.4). 
 Another challenge of non-subwavelength structures, which would become more acute for larger-aperture lenses, is that large-period gratings with only a small number of parameters per unit cell cannot easily implement the rapid variations in phase that are called for by large lenses. There are too few parameters to fit the complex intra-cell phase variation.

%diffraction_order.png

\section{Concluding remarks}
\label{sec:conclusion}

We believe that our locally periodic inverse-design approach represents a powerful extension to the ideas in previous work, allowing one to balance competing tradeoffs in wavefront design, optimize arbitrary functions of the scattered field (e.g. intensity in selected regions), evaluate parameter sensitivity, design for robustness to uncertainties, and to go beyond the regime of subwavelength structures and far-field designs. A similar max--min formulation can be used to implement a standard \emph{robust} optimization method to account for manufacturing uncertainty~\cite{MutapcicBo09, sniedovich2016statistical}. Our approximate solver remains orders of magnitude faster than optimization methods based on full Maxwell solvers, allowing it to scale to aperiodic structures hundreds or thousands of wavelengths in diameter while retaining acceptable accuracy for typical designs. We find that complex behaviors can be designed even from very simple unit cells without plasmonic resonances, and without operating in deeply subwavelength regimes.

This paper presented a proof of concept and validation of the approach, and opens up many future possibilities. We are currently working on extension to the design of 3d surfaces and vector fields, and believe such problems to be tractable with a few hours of computation (rather than the few minutes required here for 2d inverse problems). We can easily extend our inverse design from a single parameter per unit cell to multiple parameters per cell.  \edit{With a few ($\lesssim 10$) parameters, one can use a similar library-based approach via multidimensional interpolation, for which the main limitation is the number $N$ of unit-cell calculations that need to be solved beforehand in order to build the interpolation library. The simplest method is a tensor product of Chebyshev polynomials~\cite{boyd2001chebyshev}, which is practical for at most 2--3 parameters because $N$ grows exponentially with the number of parameters. Polynomial scaling of $N$ can be achieved by sparse-grid methods~\cite{Gerstner1998} \emph{or neural networks}~\cite{liu2018training,peurifoy2018nanophotonic}.  To handle hundreds or thousands of parameters per unit cell for topology optimization~\cite{yang2018freeform, lin_topology_2017, lalau-keraly_adjoint_2013,piggott_inverse_2015,piggott_fabrication-constrained_2017,sell_periodic_2017, zhan2018inverse}, the library approach must be abandoned in favor of directly solving Maxwell's equations in every metasurface unit cell for each optimization iteration (still via the locally periodic approximation). In this case, the cost is essentially independent of the number of parameters and scales linearly with the number of unit cells, which can be solved in parallel; we have successfully optimized metasurfaces with $> 1000$ parameters per unit cell in this way and are currently preparing a manuscript on those results.}
Multiple parameters per unit cell could describe more complicated surface patterns (e.g. the V-shaped antennas of \citeasnoun{yu2013flat}), but also includes the possibility of \emph{multi-layer} patterns (e.g. stacked gratings). Additional degrees of freedom could prove crucial for obtaining truly wide-bandwidth devices, coupling multiple polarizations, minimizing unwanted reflections, and so on.  

The ability to design non-subwavelength surface patterns (but still far from the $\gg \lambda$ regime of scalar diffraction theory \cite{born2013principles, o2004diffractive}) could prove useful for a variety of applications, starting with designs for short wavelengths (e.g. near UV) where subwavelength fabrication is difficult.  The additional diffracted orders of large-period structures may also become useful for near-field focusing and related design problems or for focusing a single incident beam at multiple spots.

Another interesting direction to explore would be further development of the theory of nearly periodic structures and locally periodic approximations.  In a companion work~\cite{carlos}, we develop a rigorous theory of slowly varying (nearly \emph{uniform}) structures, and show that the analogous ``locally uniform'' approximation appears as the 0th-order term in a convergent series of integral corrections. A corresponding rigorous theory of higher-order corrections to the locally periodic approximation, analogous to coupled-mode expansions for propagation \emph{through} nearly periodic media~\cite{JohnsonBi02}, along with efficient numerical methods to obtain corrections, is an important goal for the theory of metasurfaces.  A closely related problem is coupling radiation to and from guided modes by nearly-periodic surfaces, a version of which is solved in \citeasnoun{carlos}.
\edit{In another paper \cite{carlos}, we have recently shown that similar local approximations can indeed be used to compute both near fields and coupling to guided waves.}

%oe_change
\appendix
\edit{
\section*{Appendix}

To use standard high-dimensional optimization algorithms, one needs to provide an efficient computation of both the objective (cost) function and its gradient. There is a well-known technique called an \emph{adjoint method}~\cite{strang2007computational} that can be used to efficiently compute the gradient for any number of parameters with a cost comparable to evaluating the objective function at most \emph{twice}, which is commonly used in topology optimization~\cite{yang2018freeform, lin_topology_2017, lalau-keraly_adjoint_2013,piggott_inverse_2015,piggott_fabrication-constrained_2017,sell_periodic_2017, zhan2018inverse}. In the case of the two objectives presented in \secref{subsec:wavefront} and \secref{subsec:intensity}, the gradient is especially simple to evaluate as described in this Appendix.

In \secref{subsec:wavefront}, $f(p, s_0, \phi_0)=\int |s(p(x))- s_0 a(x) e^{i \phi(x) + i\phi_0}|^2  dx$, so the gradient is:
\begin{eqnarray*}
\frac{\partial f}{\partial p} &= &2\Re\left(\int \left(s(p(x))- s_0 a(x) e^{i \phi(x) + i\phi_0}\right)^* s^\prime(p(x)) dx\right)\\
\frac{\partial f}{\partial s_0} &= &-2\Re\left(\int \left(s(p(x))- s_0 a(x) e^{i \phi(x) + i\phi_0}\right)^* a(x) e^{i \phi(x) + i\phi_0} dx\right)\\
\frac{\partial f}{\partial \phi_0} &= &-2\Re\left(\int \left(s(p(x))- s_0 a(x) e^{i \phi(x) + i\phi_0}\right)^* i s_0 a(x) e^{i \phi(x) + i\phi_0} dx\right),
\end{eqnarray*}
where $\partial f / \partial p$ denotes the functional derivative~\cite{courant1957calculus} with respect to the parameter function $p(x)$ and $*$ denotes complex conjugation. Notice that the computation of the gradient requires only the evaluation of a few simple integrals, comparable to the cost of evaluating $f$.
Similarly, in \secref{subsec:intensity}, $g(p, \mathbf{x}) = |E_z(\mathbf{x})|^2 = |\int_\mathrm{y=y_0} G(\mathbf{x}, (x',0)) s(p(x')) \, dx'|^2$, and so its gradient is:
\begin{eqnarray*}
\frac{\partial f}{\partial p} &= &2\Re\left(\left(\int_\mathrm{y=y_0} G(\mathbf{x}, (x',0)) s(p(x')) \, dx'\right)^*\int_\mathrm{y=y_0} G(\mathbf{x}, (x',0)) s'(p(x'))  dx'\right) .
\end{eqnarray*}

}

\section*{Funding}

This work was supported in part by the U. S. Army Research Office through the Institute for Soldier Nanotechnologies at MIT under contract number W911NF-13-D-0001.

%%%%%%%%%%%%%%%%%%%%%%% References %%%%%%%%%%%%%%%%%%%%%%%%%
\bibliographystyle{abbrv}
\bibliography{paper1}% Produces the bibliography via BibTeX.

\end{document}